\documentclass{article}
\usepackage{amsfonts}
\usepackage{amsmath}
\usepackage{amssymb}
\usepackage{graphicx}
\begin{document}
\title{On the paradox of Hawking radiation in a  maximally extended Schwarzschild solution}
\author{George F R Ellis \\ \\
\emph{Mathematics Department and ACGC,}\\
  \emph{ University of Cape Town}}
\maketitle

\begin{abstract}
\emph{This paper considers the effect of Hawking radiation on an eternal black hole - that is. a maximally extended Schwarzschild solution. Symmetry considerations that hold independent of the details of the emission mechanism show there is an inconsistency in the claim that such a blackhole evaporates away in a finite time. In essence: because the external domain is static, there is an infinite time available for the process to take place, so whenever the evaporation process is claimed to come to completion, it should have happened earlier.
The problem is identified to lie in the claim  that the locus of emission of Hawking radiation lies just outside the globally defined event horizon. Rather, the emission domain must be mainly located inside the event horizon, so most of the Hawking radiation ends up at this singularity rather than at infinity and the black hole never evaporates away. This result supports a previous claim \cite{Ell13b} that astrophysical black holes do not evaporate.}
\end{abstract}

\section{Introduction}
\label{sec:intro}
The current dominant viewpoint as regards astrophysical black holes is that, due to Hawking radiation (\cite{Haw74}, \cite{BirDav84}), the singularity hidden behind the event horizon completely evaporates away within a finite time, so at late times there is no singularity (\cite{Haw76}, \cite{Pag76}, \cite{BirDav84}, \cite{HawPen96}).\\

The question addressed in this paper is whether this scenario is true also in the case of eternal black holes, that is, whether maximally extended Kruskal-Schwarzschild solutions completely evaporate away in a finite time due to Hawking radiation. The radiation  mechanism is the same in both cases, so the same result can be expected.  Birrell and Davies confirm that this is so: on the basis of essentially the same calculations, eternal black holes will also evaporate away in a finite time (\cite{BirDav84}, Chapter 8). Don Page (private communication) agrees, and states this as follows:
\begin{quote}
  ``\emph{I do believe that if one takes the Cauchy data for the gravitational field to be given on a Cauchy slice through the maximally extended Kruskal-Schwarzschild vacuum solution of mass $m$, and one includes quantum fields that are in a regular quantum state on this Cauchy slice that does not greatly distort the metric on this slice from the gravitational constraint equations, then the future evolution of that quantum state will include Hawking radiation that will make the black hole evaporate away in a time $\simeq 8895\, m^3$ (for this coefficient assuming $m > m_{sun}$, spherical symmetry, and emission mostly into photons and gravitons, and no other particles of rest-mass energy less than a nano-electron volt).}''
\end{quote}
I will call this \emph{The Evaporation Hypothesis}. \\

This paper shows that the Evaporation Hypothesis cannot be true, because of a paradox: this claim is inconsistent with the symmetries of the maximally extended Kruskal-Schwarzschild vacuum solution. The argument is independent of any details of the radiation emission process; it depends only on symmetry properties of the spacetime. I will call this the \emph{Hawking Radiation Paradox.}\\

I will then consider what kind of resolution of the paradox is possible. Two elements are significant here \cite{Ell13b}:
\begin{itemize}
  \item Taking in into account the back reaction effect of the Hawking radiation on the emission domain,
  \item The possibility that the emission domain is not determined by the global event horizon, but rather by locally determined Marginally Outer Trapped 3-surfaces (MOTS).
\end{itemize}
The result is still indeterminate because of the left-right symmetry and the arrow of time symmetry of the maximal Kruskal-Schwarzschild vacuum solution. A third element is required to get a clear resolution:
\begin{itemize}
  \item One can break both symmetries by imbedding the space-time in a standard cosmological context, where the evolution of the universe seta a direction of time, and cosmic blackbody background radiation, at present Cosmic Microwave Background radiation (CMB), pervades the whole universe.
\end{itemize}
The conclusion then is that most of the Hawking outgoing radiation is emitted behind the event horizon and ends up on the future singularity, rather than escaping to infinity.  Consequently blackhole evaporation does not take place in this case. This resolves the paradox.\\

The further implication is that total evaporation does not take place in the case of astrophysical black holes either, because the emission mechanism is essentially the same in both cases. Indeed I have argued in a previous paper \cite{Ell13b} that because of the three features just mentioned, astrophysical black holes do not evaporate, but rather there will always be a remnant mass left behind. The present paper supports that view.\\

An outline of the paper is as follows. Section 2 defines some key concepts. Section 3 looks at the geometry of eternal black holes, emphasizing their symmetry properties.
Section 4 presents  the current canonical semi-classical view as applied to this context.
Section 5 gives the main result: the symmetry properties of the solution contradict the Evaporation Hypothesis.
Section 6 briefly summarizes the way the elements  mentioned above lead to a new semi-classical picture where
eternal black holes result. Section 7 considers two aspects of the outcome. First, the information loss paradox is resolved by this proposal, because the black hole never radiates totally away. Second, this result supports the argument in \cite{Ell13b} that astrophysical black holes do nt radiate away.

\section{Preliminaries}
\label{sec:preliminary}
The analysis depends on the propagation of radiation and nature of trapping
surfaces for null geodesics.

\subsection{Null geodesics}
\label{sec:null_geodesics}

Radiation propagates on irrotational null geodesics with affine parameter $%
\lambda $ and tangent vector $k^{a}(\lambda )$:
\begin{equation}
k^{a}=\frac{dx^{a}}{d\lambda },k_{a}k^{a}=0,k_{;b}^{a}k^{b}=0.
\label{eq:null_geod}
\end{equation}%
The divergence $\hat{\theta}$ of a bundle of these geodesics, given by $\hat{%
\theta}=k_{;a}^{a}$, determines how the cross sectional area $A(\lambda )$
of the bundle changes:\
\begin{equation}
\frac{1}{A}\frac{dA}{d\lambda }=\frac{1}{2}\hat{\theta}.
\label{eq:null_area}
\end{equation}%
The rate of change of $\hat{\theta}$ down the null \ geodesics is given by
the null Raychaudhuri equation (\cite{HawEll73}; \cite{HawPen96}:12):
\begin{equation}
\frac{d\hat{\theta}}{d\lambda }=-\hat{\theta}^{2}-2\hat{\sigma}_{ij}\hat{%
\sigma}^{ij}-R_{ab}k^{a}k^{b}  \label{eq:null_ray}
\end{equation}%
where $\hat{\sigma}^{ij}$ is the shear of the null geodesics, and the Ricci
tensor $R_{ab}$ is determined pointwise by the Einstein field equations.

\subsection{Trapping 2 surfaces}
The gravitational field of a black hole tends to hold light in, hence it decreases the divergence $\hat{\theta}_{+}$ of the outgoing null geodesics in a spherically symmetric spacetime for any 2-sphere $%
S(r,t):=\left\{ r=const,t=const\right\} $ as $r$ tends form above to the critical value $2m$. The divergence $%
\hat{\theta}_{-}$ of the outgoing null geodesics is positive  for $r>2m.$ A marginally
trapped outer 2-surface $S_{MOTS}$ occurs when the gravitational field due
to the central mass is so large that divergence $\hat{\theta}_{+}$ of the outgoing geodesics vanishes.

\begin{quote}
\textbf{Definition: Marginally Outer Trapped 2-Surface $(S_{MOTS})$.} A
spacelike 2-sphere $S(r,t): \{\theta = const,\, \phi = const\}$  is said to be a Marginally Outer Trapped 2-Surface if the
expansion $\theta _{+}$ of the outward null normal vanishes:
\end{quote}
\begin{equation}  \label{eq:mots}
\hat{\theta}_{+}(S_{MOTS})=0.
\end{equation}%
This will happen in a Schwarzschild solution when $r_{_{MOTS}}=2m.$ For
smaller values of r, the 2-spheres $S(r,t)$ lying at coordinate values $r,t$
will be closed trapped surfaces $S_{CTS}$: that is,
\begin{equation}  \label{eq:cts}
r_{_{CTS}}<2m\Rightarrow \hat{\theta}_{+}(S_{CTS})<0.
\end{equation}%
Then as the energy condition%
\begin{equation}  \label{eq:null_energy}
R_{ab}k^{a}k^{b}\geq 0
\end{equation}%
is satisfied, by (\ref{eq:null_ray}) the outgoing null geodesics from $%
S_{CTS}$ will converge within a finite affine distance (\cite{HawPen96}:13)
and so will lie in the interior of the future of $S_{CTS}$ (\cite{HawPen96}%
:14). As these geodesics bound the causal future of the 2-sphere $S_{CTS}$,
this future will then be confined to a compact spacetime region, which
implies a spacetime singularity must occur in the future of $S_{CTS}$ (\cite%
{Pen65}; \cite{HawEll73}; \cite{HawPen96}:28).

\subsection{Trapping 3-surfaces}
\label{sec:dynamical}
When radiation is present, back reaction effects will make the horizon dynamical.
We generalise the definition of dynamical horizons by Ashtekar and Krishnan
\cite{AshKri02} by removing the restriction that it be a spacelike surface.
As these surfaces need not be dynamic, we refer to them as Marginally Trapped 3-surfaces rather than dynamic horizons Thus,

\begin{quote}
\textbf{Definition: Marginally Outer Trapped 3-Surface (MOTS).} A smooth, three-dimensional
sub-manifold $H$ in a space-time is said to be a \textit{Marginally Outer Trapped 3-Surface}%
if it is foliated by a preferred family of 2-spheres such that, on each leaf
$S$, the expansion $\theta _{(\ell )}$ of one null normal $\ell _{a}$
vanishes and the expansion $\theta _{(n)}$ of the other null normal $n_{a}$
is strictly negative.
\end{quote}

Consequently a Marginally Outer Trapped 3-Surface $H$ is a 3-manifold which is foliated by
marginally trapped 2-spheres. On this definition, such 3-surfaces can be
timelike, spacelike, or null, with rather different properties. The
essential point is that they are locally defined, and therefore are able
respond to local dynamic change. You don't need to know what is happening at
infinity in order to determine a local physical effect.\newline

The expansion $\theta _{(\ell )}$ of the outgoing null normal $\ell _{a}$
vanishes on the MOTS, and is either positive in the outside region, when it
will be called an OMOTS, or negative, when it will be called an EMOTS. The
two different kinds of MOTS surfaces have crucially different properties. An
OMOTS is the outer bound of a trapping domain; an EMOTS is the inner bound.
This difference can be expressed in terms of derivatives of the divergence $%
\hat{\theta}$ of the outgoing null geodesics. As in \cite{AshKri02},
consider outgoing null geodesics to the 2-spheres $S(v,r)$ with tangent
vector $\ell^a$, and ingoing null geodesics with tangent vector $n^a$. Then
the MOTS\ 3-surfaces, defined by $\theta _{(\ell )}=0$, are associated with
the gradient of $\theta _{(\ell )}$ in the $n^a$ direction, which determines
if the MOTS 3-surface is timelike or spacelike as follows \cite{AshKri02}
(for a proof, see \cite{Dreetal03}: Section IIB). \newline

\begin{center}
$%
\begin{array}{|c|l|l|l||}
\hline\hline
\text{MOTS 3-surface} & \partial \theta _{(\ell )}/\partial n^{a}: & \text{%
Nature:} & \text{Trapped region}:   \\ \hline\hline
\text{EMOTS} & \partial \theta _{(\ell )}/\partial n^{a}>0 & \text{timelike or null}
& \text{Inner bound}   \\ \hline
\text{OMOTS} & \partial \theta _{(\ell )}/\partial n^{a}<0 & \text{spacelike or null}
& \text{Outer bound}   \\ \hline\hline
\end{array}%
$ \\[0pt]
\end{center}

\textbf{Table 1:} \emph{Relation between trapping properties and causal
character of non-null MOTS 3-surfaces}.\newline

In the case of a null MOTS, $\partial \theta _{(\ell )}/\partial
n^{a}$ can have either sign, as can be seen from the maximally
extended Schwarzschild solution, or indeed can vanish, as can be
seen from plane wave solutions. Thus a null MOTS can be either an
EMOTS or an OMOTS.\\

We notice for future reference a key point: we have assumed here it is obvious which is the outwards and which the inwards direction. However the maximal Kruskal-Schwarzschild is degenerate, and this distinction is not always obvious. In such cases. these definitions relate to a pre-determined choice of what is the exterior and what the interior, that must be careful stated.

\subsection{Past and future trapping surfaces}
A further point is that the trapping properties of a MOTs depends on the relation to the direction of time. \\

In the above, what is outgoing and what is ingoing also depends on the choice of the direction of time. We also assume this to be given. Then there can be two kinds of MOTS 3-surfaces:
\begin{itemize}
  \item \textbf{Future directed MOTS 3-surfaces,} denoted MOTS(+), for which the definition above relates to the outgoing null geodesics in the future direction of time;
  \item \textbf{Past directed MOTS 3-surfaces,} denoted MOTS(-), for which the definition above relates to the outgoing null geodesics in the past direction of time.
\end{itemize}
Future MOTS surfaces are associated with trapping domains; they restrict where particles can go in the future. Past MOTS surfaces do not do so; the direction of time is wrong for this to happen.

\subsection{Horizons}
A \ horizon limits causal contact with the outside world. It is
necessarily a null surface and is the boundary separating events
which can escape to infinity from those that cannot. If you are
inside the horizon you cannot send
a signal to the outside world and have to end up at the singularity\\

A  MOTS surface can be a horizon, but generically is not one. Indeed it cannot be one if it is not null. However in the Kruskal-Schwarzschild solution, they are the same. This is one of the degeneracies of that solution.

\section{Classical Eternal Black Holes}
\label{sec:classical_bh}
In the case of classical general relativity, eternal black holes are characterised by existence of both past singularity and future singularities, the latter lying behind and an event horizon that hides it from the outside world (\cite{Pen65},\cite{HawEll73}, \cite{HawPen96}). The singularities are spacelike and the mass $m$ of the black hole is unchanging.

\subsection{The Kruskal-Schwarzschild solution}
\label{sec:Schw} Eternal black holes are described by the maximal
extension of the Schwarzschild spherically symmetric vacuum
solution, characterised by its mass $m$.

\subsubsection{Coordinates and metric}
The exterior part ($r>2m)$ is given in standard spherical
coordinates by \cite{HawEll73}
\begin{equation}
ds^{2}=-(1-\frac{2m}{r})dt^{2}+\frac{dr^{2}}{(1-\frac{2m}{r})}+r^{2}(d\theta
^{2}+\sin ^{2}\theta d\phi ^{2}) .  \label{eq:Sch}
\end{equation}%
The timelike world line $\{r=\text{const},\theta =\text{const%
},\phi =\text{const}\}$ for $r>2m$ are symmetry orbits: every point on these curves is physically equivalent to every other one. Proper time along them is given by
\begin{equation}
\Delta\tau:=\tau_1-\tau_2 = \left(\sqrt{1-\frac{2m}{r}}\right) (t_1-t_2).  \label{eq:propertime}
\end{equation}%

\subsubsection{Extension to $r<2m$}
The metric (\ref{eq:Sch})\ is singular at $r=2m$, but there exist regular coordinates across this null surface \cite{HawEll73}. Define
\begin{equation}  \label{eq:Schw_rstar}
r^{\ast }=\int \frac{dr}{1-2m/r}=r+2m\log (r-2m);
\end{equation}%
then $v:=t-r^{\ast }$ is an advanced null coordinate and $w:=t+r^{\ast }$ is a retarded null coordinate.
By using coordinates $(r,v)$ one attains an extension across the horizon $r=2m$, joining the exterior and interior regions; using coordinates $(r,w)$ one attains an extension across the horizon $r=2m$ joining the exterior to a different interior region. One has two add a further exterior domain to get the maximal extension (see \cite{HawEll73} for  details).\\

On conformally rescaling these coordinates, the metric (\ref{eq:Sch}) can be rewritten in a double-null form which is regular for all $r>0.$ This can then be presented in terms of a Penrose-Carter conformal diagram
shown in Figure 1 (\cite{HawPen96}:45).
\begin{figure}[tbp]
\includegraphics[width=7in]{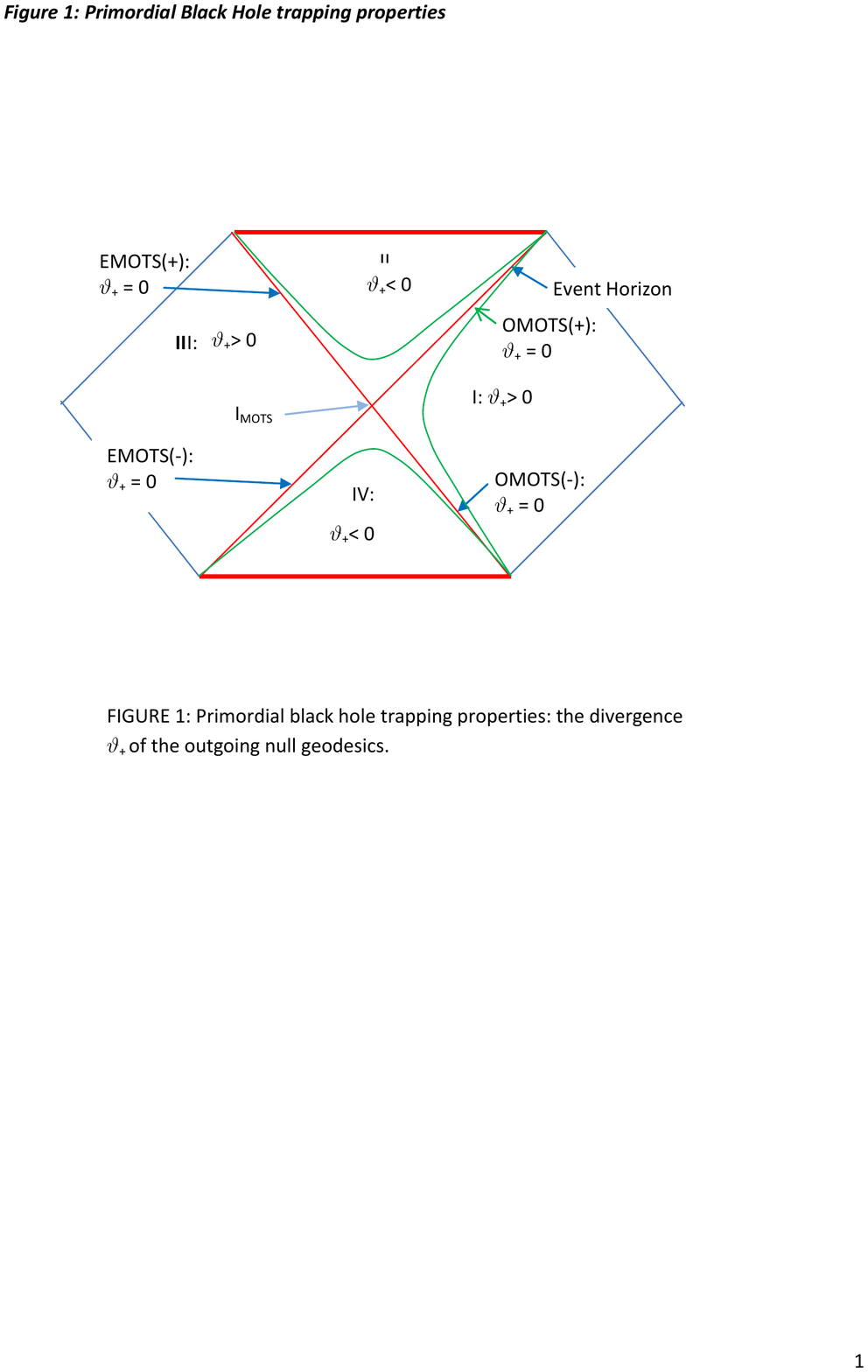}
\caption{Figure 1}
\label{Fig1}
\end{figure}
\subsection{Domains}
The solution has four domains labelled I-IV, connected across null horizons, It is static for $r>2m$ (as $%
\partial/\partial t$ is a timelike Killing vector field), and hence Domains I and III region are eternal (they do not come to an end at any finite positive or negative value of the time parameter $t$). The solution is dynamic for $r<2m$, which occurs in Domains II and IV. The green lines mark surfaces of constant $r$, timelike in domains I and III, spacelike in domains II and IV.
\begin{enumerate}
\item \textbf{Domain I:} An outer vacuum spacetime, bounded on the outer
sides by past and future null infinity, and on the inner side by an OMOTS(+) surface $r=2m$ in the future and an OMOTS(-) surface $r=2m$ in the past. It is static and hence is eternal.
\item \textbf{Domain II}:\ The vacuum spacetime inside the event horizon. It is a spatially homogeneous but time
evolving part of a Schwarzschild solution. It is comprised of closed trapped surfaces, which imply existence of the future spatial singularity that bounds Domain II to the future. It is bounded in the past by two OMOTS(+) surfaces $r=2m$.
\item \textbf{Domain III:} A mirror image of Domain I, with the infinite boundaries on the left rather than the right  It is again static.
\item \textbf{Domain IV}: A mirror image of Domain II, but with the direction of time reversed. It is comprised of time reversed closed trapped surfaces, which imply existence of the spatial singularity that bounds Domain IV to the past. It is again dynamic.
\end{enumerate}

\subsection{Boundaries}
The boundaries between the domains are,
\begin{itemize}
\item \textbf{B12:} Between I and II, the event horizon given by $r_{H}=2m$, which is a Killing horizon and the locus of a family of $S_{MOTS}$ 2-spheres, i.e. each 2-sphere in the event horizon is a $S_{MOTS}$ with $%
\hat{\theta}_+=0$. Thus it is an OMOTS(+) surface. Objects inside $r_{H}$ are trapped, because $r=r_{H}$ is a null surface.  Thus it is an EMOTS(+) surface. The event horizon hides events in the interior from the exterior domain I.
\item \textbf{B23:} Between II and III, the mirror-symmetric event horizon again given by $r_{H}=2m$, which   hides the interior from the exterior domain III.
\item \textbf{B34:} Between III and IV, a time-symmetric version of B23.   Thus it is an EMOTS(-) surface.
\item \textbf{B41:} Between IV and I, a time symmetric version of B12.  Thus it is an OMOTS(-) surface.
\end{itemize}
The inmost trapped 2-surface ($I_{MOTS}$) is the 2-sphere at the
join of domains I, II, III, and IV. This is a bifurcate Killing
horizon \cite{Boy69}. The area of the surfaces $r=2m$ is in each case constant along the outgoing null geodesics for all $\lambda $ because equation (%
\ref{eq:null_ray})\ is satisfied with $\hat{\theta}=0,\ \hat{\sigma}_{ij}%
\hat{\sigma}^{ij}=0,R_{ab}k^{a}k^{b}=0.$ \\

\noindent Figure 1 shows the sign of the outgoing null geodesic expansion $\hat{\theta}_+$ in each domain, and so how the MOTS surfaces mark the changes in sign in this expansion.

\subsection{Symmetries}
The solution has many symmetries which will play a key role in what follows
\begin{figure}[tbp]
\includegraphics[width=7in]{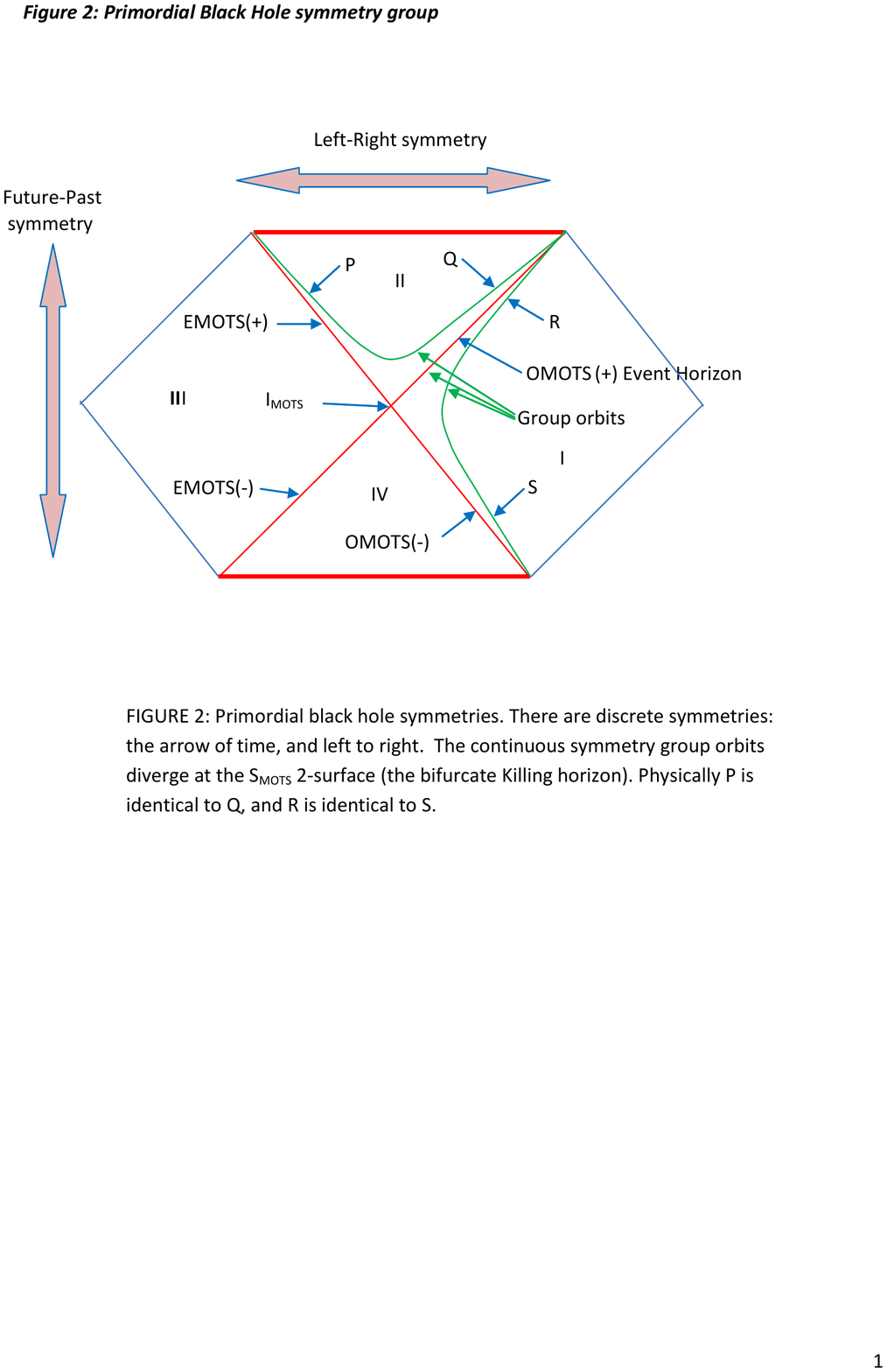}
\caption{Figure 2}
\label{Fig2}
\end{figure}

\subsubsection{Left-Right symmetry}\label{sec:space_sym}
There is a left-right symmetry in the solution: change the obvious conformal coordinates $(t,x)$ in the figure centred on the $I_{MOTS}$ 2-sphere to $(t,-x)$ and the solution is invariant. Therefore Regions I and III are identical, but with the spatial direction reversed; while Regions II and IV are individually spatially symmetric.

\subsubsection{Direction of time symmetry}\label{sec:time_sym}
There is a time symmetry in the solution: change the conformal coordinates $(t,x)$  to $(-t,x)$, and the solution is invariant. Therefore Regions I and III are individually time symmetric, while Regions II and IV are symmetric to each other  but with the time direction reversed.

\subsubsection{Boost group symmetry}\label{sec:boost}
The solution is invariant under the Lorentz boost group, with the $I_{MOTS}$ 2-sphere the set of fixed points \cite{Boy69}. The surfaces of constant $r$ are the orbits of this group: timelike in Domains I and III (hence static there), spacelike in Domains II and IV (hence spatially homogeneous there), null on the surfaces $r=2m$.\\

Consequently, events P and Q are equivalent in all respects, because the symmetry group moves them into each other. No physical feature can distinguish them. The same is true of events R and S. It is noteworthy that the Killing orbits bifurcate at the $I_{MOTS}$ 2-sphere, hence events Q and that are close to each other but on opposite side s of the horizon $4r=2m$ are related by the group symmetry to the very distant points P, S respectively. This is the divergence that underlies key aspects of the behaviour we discuss below.

\subsection{Outcomes}

A spacelike singularity occurs both in the future and past as $r \rightarrow
0$. The Weyl tensor diverges there (the Ricci tensor is zero).
Specifically, the Kretschman scalar is
\begin{equation}  \label{eq:Kretsch}
K\ =C_{abcd}C^{abcd}=\alpha \frac{m^{2}}{r^{6}}
\end{equation}%
where $\alpha =48G^{2}/c^{4}$ \cite{Hen99}, so this diverges as $%
r\rightarrow 0.$ It is the spatial inhomogeneity of the solution
that generates this singularity in the conformal structure of
spacetime.\newline

As seen from the outside, the mass of the star never alters; it is always
equal to the initial value $m_0$:
\begin{equation}  \label{eq:mass_const}
m = m_0 = \mathit{const}.
\end{equation}
This will of course not be the same if matter falls into the blackhole,
hereby increasing it is mass; then the horizon is a dynamic horizon \cite%
{AshKri02} and the laws of black hole thermodynamics \cite{BarCarHaw73} come
into play to characterise the resulting changes. However we do not consider
those processes here.

\section{Semiclassical gravity: The standard View}
\label{sec:semiclassical}
This is all altered when one takes quantum field theory into account, so
that Hawking radiation results in mass loss and (\ref{eq:mass_const}) is no
longer true.\\
Three new effects come into play, and dramatically alter the picture (Figure 3).
\begin{figure}[tbp]
\includegraphics[width=7in]{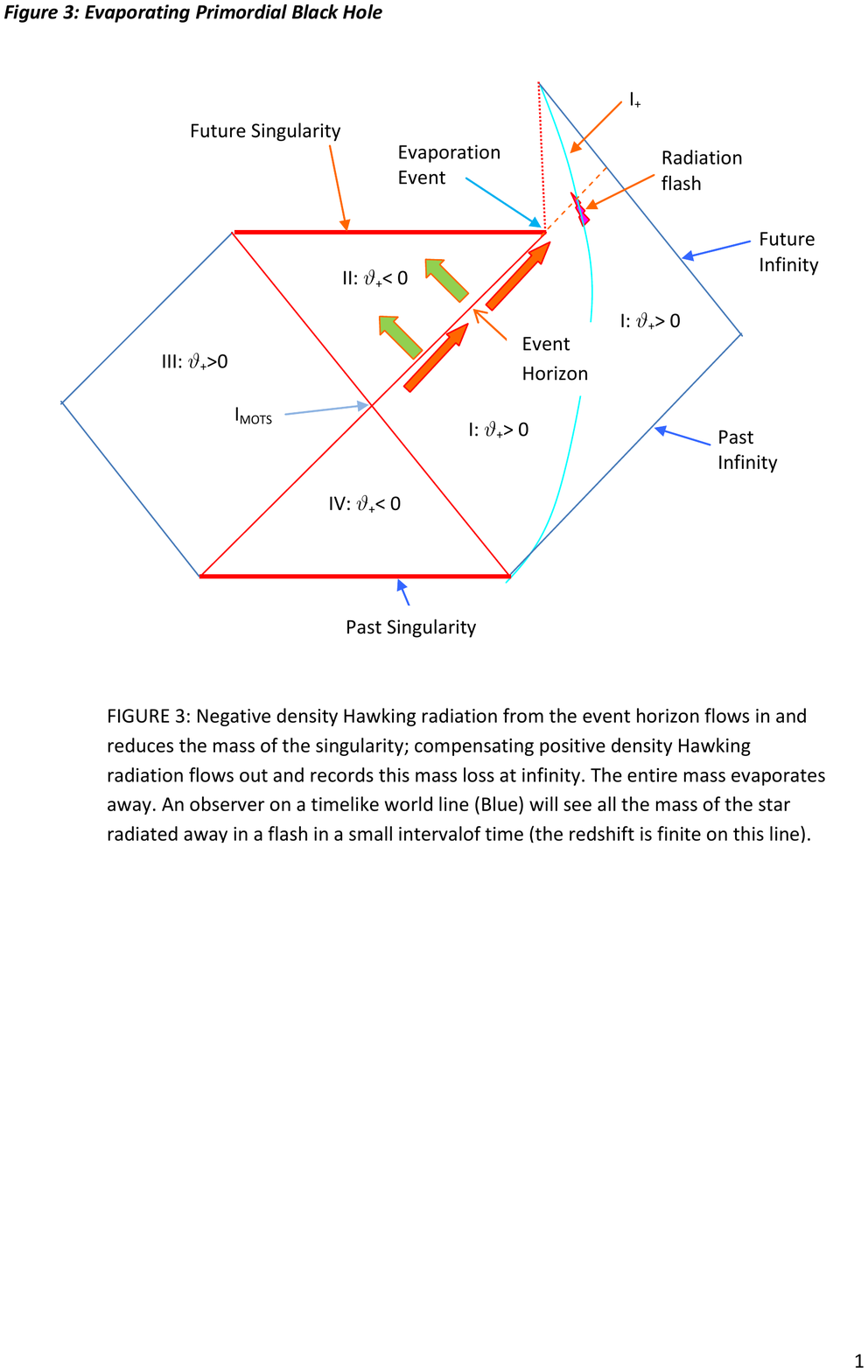}
\caption{Figure 3} \label{Fig3}
\end{figure}
\newline
The effects are,\\

\textbf{Item 1}: Quantum field theory fluctuations leads to production of
black body Hawking radiation at the event horizon (\cite{Haw73}; \cite%
{BirDav84}, \cite{HawPen96}:43), with temperature determined by the mass of
the central body:
\begin{equation}
T_{BH}=\frac{1}{8\pi m}.  \label{Eq:T}
\end{equation}
This result has been calculated in many ways, and is independent of the
gravitational field equations \cite{Vis01}. In order to conserve energy, outgoing positive density Hawking radiation emitted just outside the event horizon is balanced by negative density ingoing Hawking radiation.   This outgoing positive density black body radiation escapes to infinity. Consequently in thermodynamic terms, the black hole acts as a black body with temperature $T_{BH}$ given by (\ref{Eq:T}) and entropy
\begin{equation}  \label{Eq:S}
S_{BH}=4\pi m^{2}=\frac{1}{4}A.
\end{equation}
This latter result is dependent on the gravitational field equations \cite{Vis01}. \\

\textbf{Item 2}: Because radiation is being emitted, energy conservation
shows there must be a corresponding mass loss by the black hole:
\begin{equation}
dm/d\tau <0,  \label{eq:dmbydu}
\end{equation}%
where $\tau $ is give by (\ref{eq:propertime}) in the outer domain, so $m$
decreases with $\tau $ and therefore $T_{BH}$\ increases as the radiation
process continues. One can think of the outcome as like a Schwarzschild
solution with ever decreasing mass $m(\tau )$.\newline

\textbf{Item 3}: The singularity that has formed at the centre eventually
pops out of existence (\cite{Haw74}; \cite{BirDav84}; \cite{HawPen96}) because the Hawking process inevitably
carries all the mass away to infinity. This happens in a finite time,
because as the mass decreases the radiation loss process speeds up. The
power $P$ radiated is proportional to $1/m^{2}$; using the usual mass-energy
equivalence,
\begin{equation}
P=-c^{2}dm/d\tau ,  \label{eq:power}
\end{equation}%
the evaporation time $t_{evap}$ is proportional to $m_{0}^{3}$ (\cite{Haw76}%
, \cite{Pag76}). Putting in the numbers, this lifetime will be of the order
\begin{equation}
t_{evap}\simeq 10^{71}(m/m_{\odot })^{3}secs.  \label{eq:lifetime}
\end{equation}
The crucial new feature is that the effect of the Hawking radiation
is to eventually make both the central mass and singularity vanish in a finite time.
The result is that the outer
domain eventually loses the event horizon and again has a regular
centre.

\section{The Paradox}
The key point is that the exterior domain is static. Any time that evaporation starts and
vaporizes the mass, it should have started earlier, because there is
an infinite preceding time when the process can take place. When
ever it gets completed, it should have happened before.\\

    In detail: Given
    that the evaporation process takes a finite time $\tau_{evap}(m_0)$
    determined by the initial mass $m$, whenever the evaporation
    ends, say time $\tau_{final}$, it must have started at time
    $\tau_{1}=\tau_{final}-\tau_{evap}(m_0)$ when the radiation achieve some threshold
    value we regard as the starting value of the process. But conditions then
    were identical to an even earlier time
    $\tau_{2}=\tau_{1}-\tau_{evap}(m_0)$. So it should have
    evaporated away by time $\tau_{1}$ rather than starting at time
    $\tau_{1}$. This argument repeats: whatever time is supposed to
    be the end time, the process should have ended already before it
    started. There is no finite time at which it can start. Hence
    there is no determinate time when it can end.\\

\begin{figure}[tbp]
\includegraphics[width=7in]{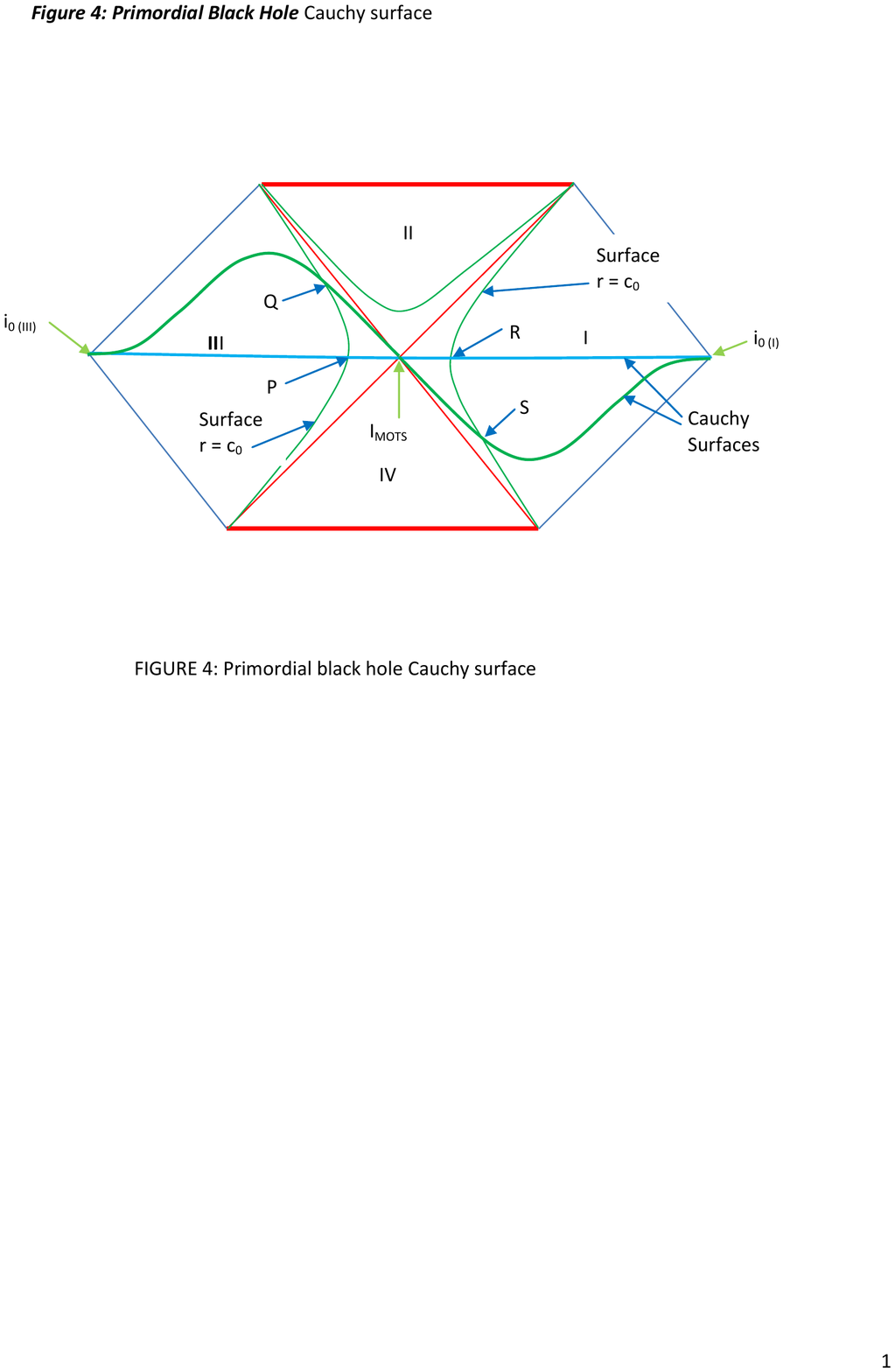}
\caption{Figure 4} \label{Fig4}
\end{figure}
The argument can be made more formal in terms of Cauchy surfaces, as mentioned in Page's formulation of the  Evaporation Hypothesis  (Section \ref{sec:intro}). The situation is shown in Figure 4.\\

Consider a star of mass $m > m_{sun}$ and a timelike Killing vector orbit $r = c_0 >2m$ in Domain I, as shown. Define $\Delta \tau = 8895\, m^3$. Let the event $R$ on the world line $W$ be defined by $r_W = c_0 =2m(1+\delta)$, $\delta>0$, $|\delta \ll 1|$. Then  $W$ is close to the event horizon. Let the event $S$ also be on the world line $W$, but a proper time $\Delta \tau$ in the past from $R$ along $W$ as given by (\ref{eq:propertime}). \\

Consider first a Cauchy Surface $CS_1$ that starts at spatial infinity $i_{0 (I)}$, goes through the event $R$ on the world line $W$ in Domain I, then through the central $I_{MOTS}$ surface, then through the event $Q$ on the mirror world line $r = c_0$ in Domain III, and ends at spatial infinity $i_{0 (III)}$.\\

Consider a second Cauchy Surface $CS_2$ that is determined from $CS_1$ by an action of the boost symmetry group discussed in Section \ref{sec:boost}, with the group element chosen so that $CS_2$ passes through the event $S$. Then $CS_2$ starts at spatial infinity $i_{0 (I)}$, goes through the event S on the worldline $W$ in Domain I, then through the central $I_{MOTS}$ surface (the fixed surface of the group), then through the event Q on the other surface $r = c_0$ in Domain III, and ends at spatial infinity $i_{0 (III)}$. The form of the surface is because $i_{0 (I)}$, $I_{MOTS}$, and $i_{0 (III)}$ are fixed points of the group; and it acts downwards in Domain I but upwards in Domain III.\\

If we set initial data as specified in the  Evaporation Hypothesis (Section \ref{sec:intro}) on the Cauchy Surface $CS_1$, the black hole will evaporate a time $\Delta \tau$ in the future of $R$ as measured along the world line $W$. But there is nothing special about this Cauchy surface; its choice was arbitrary. We could have set the same data on the physically identical Cauchy surface $CS_2$ through the event $S$. Then the black hole would have evaporated by the time of event $R$. And so on: we could have set data in an even earlier Cauchy surface $CS_3$ so that the black should have evaporated before $CS_2$.Whatever Cauchy surface we choose, there is an earlier one such that the black hole will be gone before we set that data. Continuing back in time, given the eternity allowed between the event $S$ and the start of the spacetime at $\tau \rightarrow - \infty$, the black hole must have evaporated before any finite time which we can consider to be the start of the Hawking evaporation process (which is claimed to happen in a finite time).\\

Thus we have deduced
\begin{quote}
    \textbf{The Hawking Radiation Paradox}: \emph{the emission and evaporation picture is inconsistent
    with the symmetry of the extended Schwarzschild solution.}
\end{quote}
The inconsistency is quite independent of the details of the process; it only hinges on the claim that the process ends a finite time after a chosen starting time, together with the static nature of the maximal Kruskal-Schwarzschild solution exterior Domain I.

\section{Elements of a resolution}
\label{sec:revised}
The basic issue is where Hawking radiation emission takes place, and hence where it goes to.\\

One can calculate the vacuum expectation value of the energy
momentum tensor of the relevant field. This has been done in the
case of a two-dimensional model of the collapse of a shell of matter
by Davies, Fulling and Unruh \cite{DavFulUnr76}. They find,
\begin{quote}
``\emph{The flux of energy is given by two components. Near infinity it is
dominated by an outward null flux of energy. Near the horizon, however, it
is a flux of negative energy going into the horizon of the black hole. }.''
\end{quote}
Thus we can think of there being an emission domain just outside the
horizon, which emits positive energy radiation going outwards and negative
radiation going inwards.\\

 But the issue is where does this process takes place: which horizon is it?

\subsection{The basic argument}\label{sec:basic1}
Figure 4 gives the outlines of argument that proposes a resolution.
\begin{figure}[tbp]
\includegraphics[width=7in]{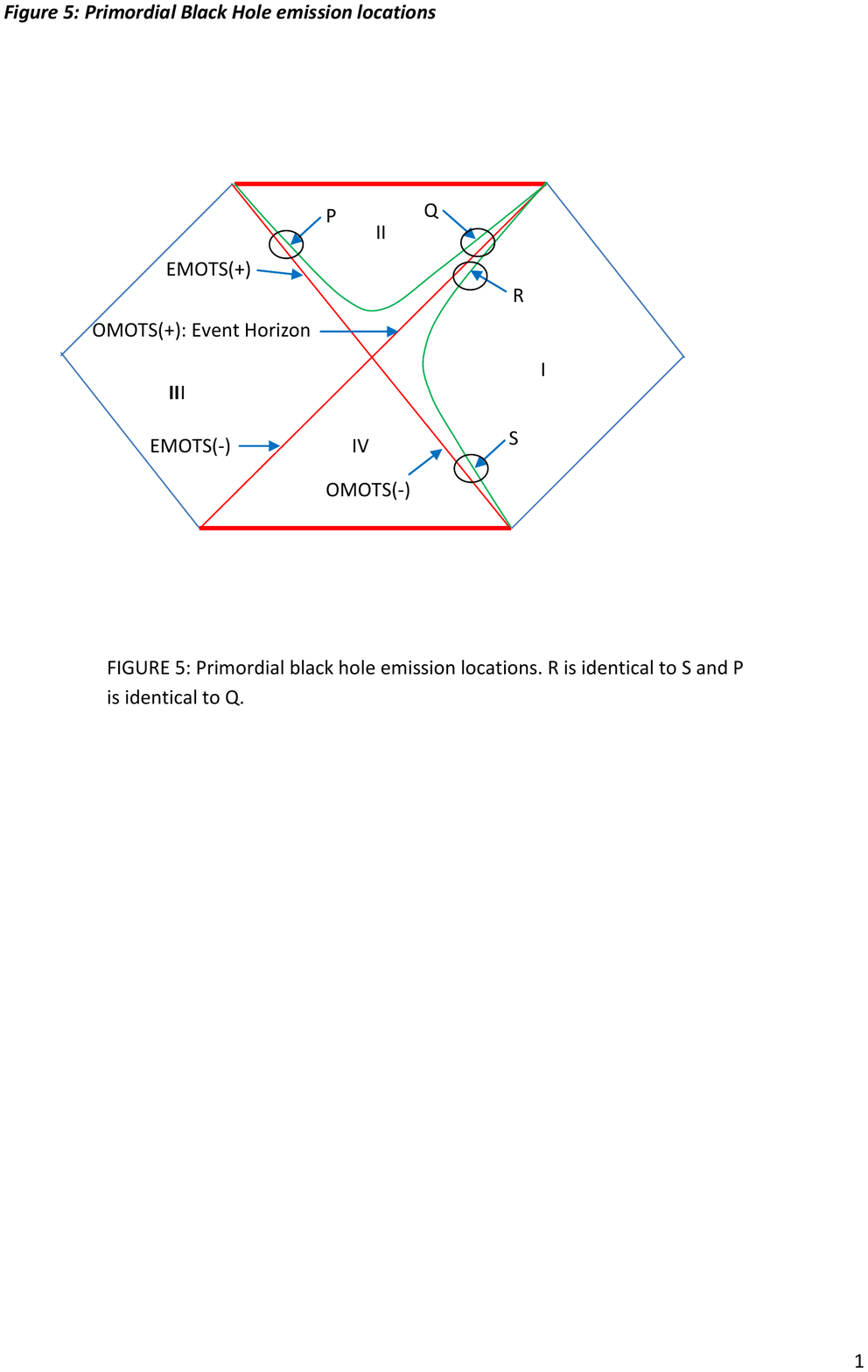}
\caption{Figure 5} \label{Fig5}
\end{figure}
The key point was made in section \ref{sec:boost}: because of the boost symmetry group, conditions at $Q$ and $R$ are identical; also conditions at $P$ and $Q$ are identical. \\

Now it is usually supposed that Hawking radiation is emitted just outside the event horizon, that is, at  a point like $R$.  But conditions at this point just outside the OMOTS(+) are physically identical
to conditions at point $S$ just outside the OMOTS(-), by this symmetry group. However considering a small neighbourhood of $S$, we do not expect radiation for here, because it is close to an anti-trapping surface (an OMOTS(-)) rather than being close to a trapping surface. But as conditions here are identical to those close to $R$, we should not expect radiation from close to $R$. The latter cannot radiate because the former does not (it is outside a non-trapping surface).\\

By contrast, consider the point $Q$ just inside the EMOTS(+) surface. Its physical conditions are  identical
to those at the $P$ point just inside the OMOTS(+) surface. Conditions at $P$ are conducive to emission of Hawking radiation (cf. the discussion below), thus the domain near Q  can radiate because the its conditions are identical to those at $P$. We can expect radiation from near $P$ and $Q$ rather than from near $R$ and $S$.\\

In summary,
\begin{itemize}
  \item Domains just outside the OMOTS(+) surface are also just outside the OMOTS(-) surface and so should not radiate.
  \item Domains just outside the EMOTS(+) surface are also just inside the OMOTS(+) surface and are the ones that radiate.
\end{itemize}
If this is so, the locus of radiation emission is Domain II rather than Domain I. That means both ingoing and outgoing Hawking radiation ends up at the future singularity rather than at infinity.
Hence, just as claimed in \cite{Ell13b} for the case of astrophysical black holes, the black hole mass will not radiate away. This resolves the paradox; there is no special time for the start of the evaporation process, because  it never takes place.\\

Developing this into a full revised view has three elements: taking into account backreaction effects (Section \ref{sec:Back}), reconsidering the locus of radiation emission (Section \ref{sec:Emission region}), and breaking the degeneracies of the solution by imbedding it in a cosmological context (Section \ref{sec:cosm}). These will only be briefly considered here, because all these issues are discussed in depth in \cite{Ell13b}. The argument is a straightforward generalisation of what is presented there.

\subsection{Back reaction effects}
\label{sec:Back}
The effect of the Hawking radiation is to displace both the OMOTS(+)
and the EMOTS(+) trapping surfaces.

\subsubsection{The OMOTS(+) Surface}
Because of the null Raychaudhuri equation (\ref{eq:null_ray}), the
outgoing positive density Hawking radiation causes focussing of null geodesics, so any
outgoing null geodesics that start at an initial affine parameter
$\lambda_1$ at a $S_{MOTS}$ 2-surface where $\hat{\theta}_+
= 0$ will converge to a conjugate point at finite affine parameter value $%
\lambda_2\,>\,\lambda_1$. Hence a marginally trapped
$S_{MOTS}(\lambda_1)$ in the OMOTS 3-surface will be mapped by the
outgoing null geodesics to
trapped 2-spheres $S_{CTS}(\lambda)$ with $\hat{\theta}(\lambda) < 0$ for $%
\lambda_1\,<\,\lambda\,<\,\lambda_2$. These 2-spheres, whose future
necessarily ends on a singularity (because they are trapped!), will
therefore lie inside the OMOTS surface; hence the OMOTS surface has
moved out relative to the null geodesics through
$S_{MOTS}(\lambda_1)$. \newline

Consequently, the effect of the radiation is to change the OMOTS
3-surface from null to spacelike.  It now lies outside the initial
null surface $r=2m$ generated by the geodesics through the $I_{MOTS}$
2-sphere. It bounds the domain of trapped surfaces, and so it bounds
the events whose future ends up at the singularity. Thus the OMOTS
in fact defines the extent of the future singularity.\newline

The further important consequence is that the event horizon is no
longer where it used to be. It used to be generated by the outgoing
null geodesics starting at the IMOTS 2-sphere where $r=2m$, and
reaching the future singularity at the point $P_1$. That initial
null surface is now trapped. \newline

\textbf{In summary:} the effect of the radiation is to change the
OMOTS 3-surface from being a null surface that coincides with the
event horizon, to being a spacelike surface that
\begin{enumerate}
\item lies outside the initial null surface,
\item determines the extent of the future spacelike singularity, and
\item thus determines the location of the event horizon.
\end{enumerate}

\subsubsection{The EMOTS(+) surface}
In a dual interaction, the ingoing negative density Hawking radiation defocuses outgoing null geodesics, and so lifts the EMOTS surface off the null surface $r=2m$ and makes it timelike, lying inside Domain II. Details are in \cite{Ell13b}.

\subsection{The Emission region}
\label{sec:Emission region}
The central issue that determines the outcome is, Where is the outgoing radiation emitted? \\

The usual proposal (e.g. \cite{Pag76} is that it is emitted just outside the globally determined event horizon. However there is a counter view (\cite{Haj87}, \cite{ParWil00}, \cite{Vis01}, \cite{Nie08},  \cite{Cli08}, \cite{Pel09}, \cite{ParPqad09}) that it is rather emitted just outside a locally determined MOTS surface.
In that case, this would radiation emission not occur if the MOTS
surface were spacelike:
\begin{itemize}
\item If we use the tunneling description \cite{ParWil00}, the MOTS must be
timelike else there is no surface to tunnel through. The very concept of
tunneling depends on the implicit assumption that the trapping surface
causing the emission must be timelike so that two sides (``inside'' and
``outside'') can be defined.
\item The concept of scattering involved in use of Bogoliubov
transformations \cite{Pel09} assumes scattering is off a timelike world tube.
\end{itemize}
Hence for the Hawking radiation emission to occur, the trapping
surface must be a timelike MOTS, as in the following Table.\\

$%
\begin{array}{|l|l|l|l|l|l|l|}
\hline\hline
\text{Horizon} & \theta _{(\ell )} & \partial \theta _{(\ell )}/\partial
n^{a} & \text{Nature:} & \text{Radiation?} & Mass &  \\ \hline\hline
\text{EMOTS} & \theta _{(\ell )}=0 & \partial \theta _{(\ell )}/\partial
n^{a}>0 & \text{timelike} & \text{emits radiation} & m_{in} &  \\ \hline
\text{OMOTS} & \theta _{(\ell )}=0 & \partial \theta _{(\ell )}/\partial
n^{a}<0 & \text{spacelike} & \text{emits no radiation} & m_{out} &  \\
\hline\hline
\end{array}%
$\newline
\newline

\textbf{Table 2}: \emph{Outgoing null geodesic divergences according to
domain}.\newline

\noindent This leads to the alternative view proposed in \cite{Ell13a}:
\begin{quote}
\textbf{\ Main Hypothesis}: The source of outgoing Hawking radiation
is neither near the event horizon, nor outside the spacelike outer trapping (OMOTS) surface
surface: its location is outside the inner (EMOTS) timelike trapping surface, and hence  the inside the trapping domain.
\end{quote}
The consequence is that, in agreement with the argument above (Section \ref{sec:basic1}), the domain of emission is Domain II rather than Domain I. Most of the radiation is trapped and does not get to infinity.

\subsection{Breaking the symmetries}\label{sec:cosm}
However there remains a double symmetry paradox.

\subsubsection{The spatial paradox}

\begin{figure}[tbp]
\includegraphics[width=7in]{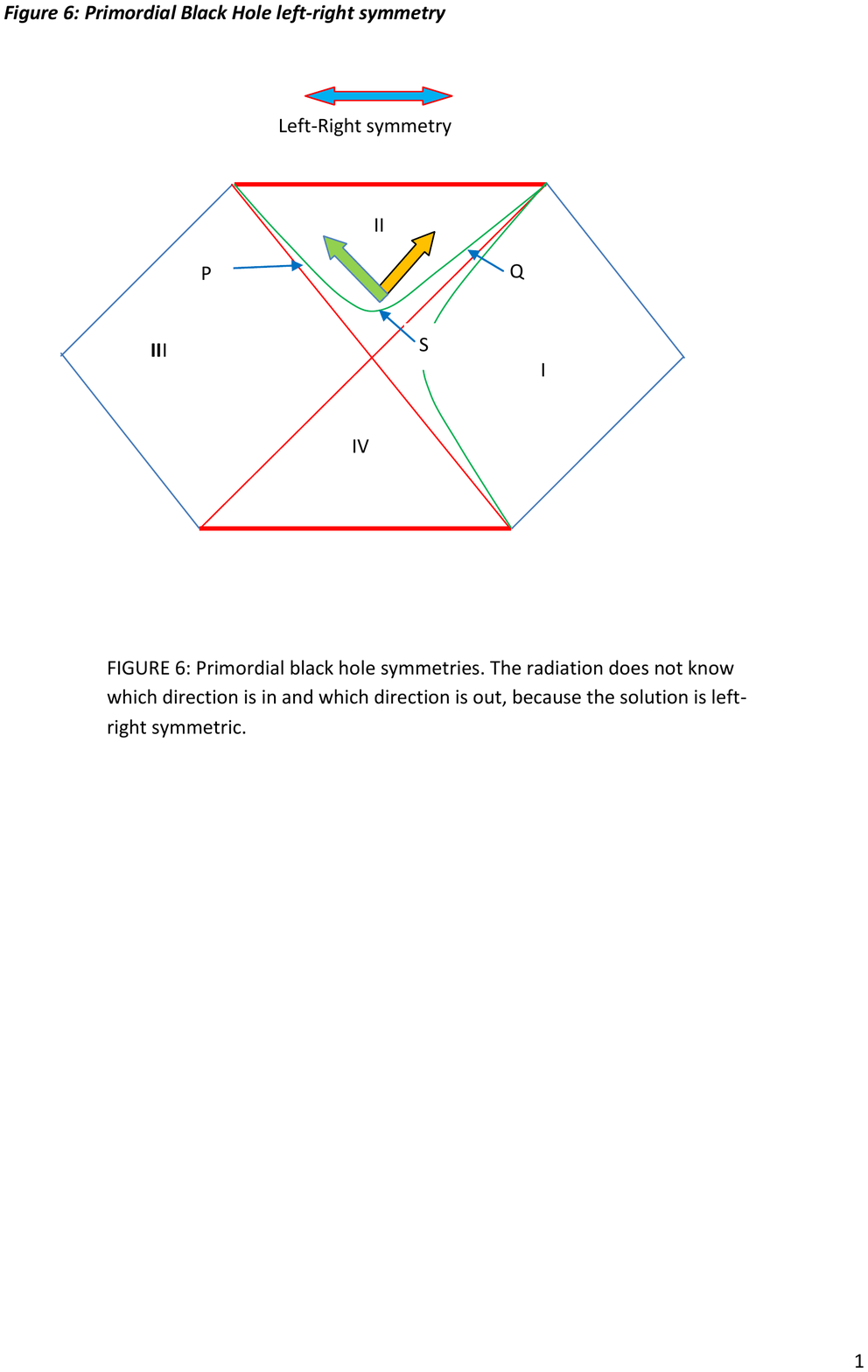}
\caption{Figure 6} \label{Fig6}
\end{figure}
In the above, it was assumed we had a good definition of in and out. But because of the
spatial symmetry discussed in Section \ref{sec:space_sym}, this is completely ambiguous, see Figure 5. \\

Looking at the geometry as relates to $P$ or $Q$, it seems quite clear what is in and what is out. But this is an
illusion. Both points are physically fully equivalent to $S$, where the distinction is fully ambiguous. How does the physics know which is the in direction, where the negative density Hawking radiation should go, and which is the out direction, where the positive density Hawking radiation should go? \\

Given the ambiguity, does Hawking radiation obey this symmetry and send positive and negative energy density radiation equally to the left and the right? if this is the case, the outcome will be different than usually envisaged. If not, why not? Is one direction chosen randomly by some quantum event? If so, why is the direction the same everywhere?\\

A resolution can be obtained as in \cite{Ell13b}: introducing a cosmological context, imbedding Domain I (but not Domain II) in an expanding universe (\cite{EllMaaMac12}, \cite{PetUza09}) will break the symmetry and determine what is in and what is out. The incoming CMB radiation, effectively emitted at a finite infinity surrounding the black hole (\cite{Ell84}, \cite{Ell02}), will then reinforce the effect of Hawking radiation on the OMOTS surface discussed in Section \ref{sec:Back}.

\subsubsection{The time paradox}
There is an equal issue as regards the time symmetry of the solution described in Section \ref{sec:time_sym}. How does the radiation know which is the future and which is the past?\\

As in \cite{Ell13b} introducing a cosmological context will also break this symmetry. It defines
what is the future and what is the past \cite{Ell13a}.
\subsection{The Outcome}
\label{sec:main}
We now have all the pieces needed to determine where Hawking radiation
emission takes place in a maximal Kruskal-Schwarzschild blackhole setting.
The broad nature of the resulting spacetime is shown in Figure 6, indicating the sign of the outgoing
null geodesic expansion $\hat{\theta}_+$. The changes in sign of $\hat{\theta%
}_+$ determine the location of the MOTS surfaces, and hence characterises
where radiation will be emitted.
\begin{figure}[tbp]
\includegraphics[width=7in]{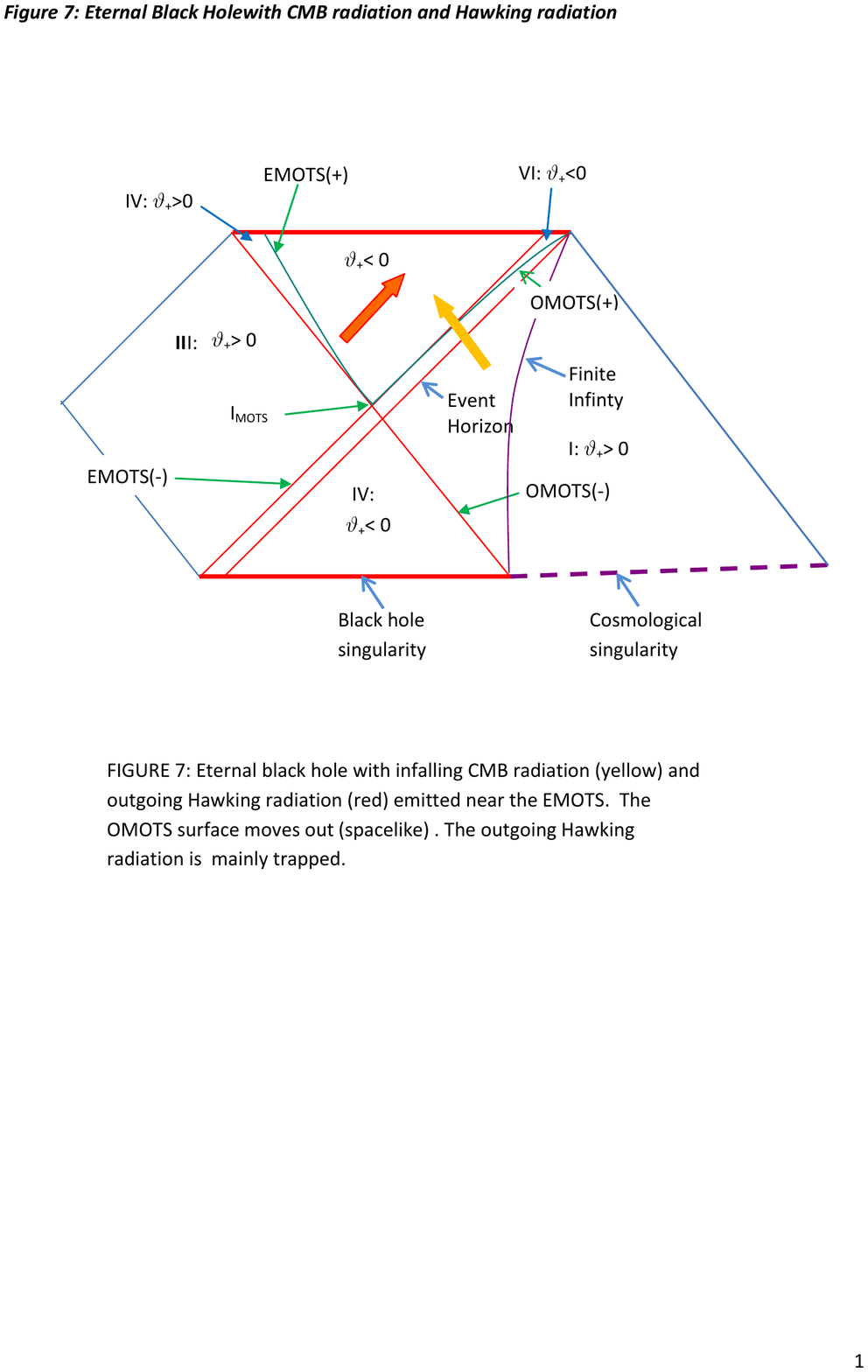}
\caption{Figure 7} \label{Fig7}
\end{figure}
An immediate consequence is

\begin{quote}
\textbf{Conclusion}: Most of the Hawking radiation emitted does not end up at infinity: it ends up on the future
singularity. Consequently the singularity does not evaporate away because of
outgoing Hawking radiation; that radiation mainly ends up on the singularity, and
so does not carry mass or energy away to the exterior
\end{quote}

This in full accord with the analysis of the previous sections. However it must be emphasized that the analysis of this section is based on the particle tunneling model of Parikh and Wilczek \cite{ParWil00}; Paul Davies emphasizes one needs to calculate the stress tensor of the radiation (\cite{Dav76}, \cite{DavFulUnr76}) to confirm the result. This is still to be done; Clifton \cite{Cli08} has shown there is no problem in calculating the stress tensor in the case of emission by  a local trapping surface rather than a global event horizon. However the arguments of the previous sections are independent of any specific emission mechanism, and give strong support to the picture presented here.

\section{Implications}
\label{sec:option2_Implications}
This viewpoint has obvious implications for the current debates on
the information loss paradox.\newline

Is there information loss? (\cite{Haw76}, \cite{Haw05}, \cite{Mat12}, \cite%
{Gid13}). Yes indeed (\cite{HawPen96}: 59, 63). Gravity is a non-linear
theory with spacetime singularities:\ information falling into them is
inevitably lost. Any matter or information falling in through the event
horizon disappears into the permanent spacelike singularity, which does not
go away and can act as a sink for an arbitrary number of microstates. As no
radiation is emitted outwards from the singularity to infinity (because it
is spacelike), no information can be carried out of the black hole from the
singularity by any such radiation. Because there is a permanent relic, just
as in the case Maldacena and Horowitz \cite{HorMal04}, there is no
information loss paradox.\newline

Conclusion: Infalling matter and information falls into the singularity and
is destroyed there. It cannot be re-emitted by Hawking radiation from there
as no Hawking radiation from there reaches infinity. Microstates are
swallowed up by the singularity. From the outside, the black hole acts as an
absorbing element, so scattering off it should not be expected to be unitary
(energy will not be conserved at the event horizon). Additionally, the
astrophysical context indicated here (specifically, the presence of the CMB
radiation) causes rapid decoherence, so entanglement across the horizon \cite{Mat12} is rapidly lost
\cite{Sch07} and associated problems will therefore
dematerialize. But in any case entanglement is across the EMOTS surface, not the event horizon. Hence the particles can remain entangled in a unitary way until they hit the future singularity.\\

If the account presented in this paper is correct, eternal black holes do not evaporate away; and this will also apply also to the case of astrophysical black holes, because the evaporation mechanism is the same, even though the context is different. Thus this paper supports the arguments in \cite{Ell13b}.
What these discussion show is that if \emph{any} of the
Hawking radiation falls into the singularity, it is unlikely that it can
then evaporate away; and then the broad picture presented here will be
correct. And even if one changes many details, it is unlikely that \emph{all} the Hawking radiation can
avoid a singular fate of this kind. Then the main result will be vindicated.
\newline

\bigskip

\noindent

\textbf{Acknowledgements:}\newline

I thank Don Page for challenging my views on these topics,and Paul
Davies, Tim Clifton, and Samir Mathur for helpful comments.\newline

I acknowledge the financial support of the University of Cape Town Research
Committee and the National Research Foundation (South Africa). \newpage



\begin{thebibliography}{99}

\bibitem{AshKri02} A Ashtekar and B Krishnan (2002)\ \textquotedblleft
Dynamical Horizons: Energy, Angular Momentum, Fluxes and Balance
Laws\textquotedblright\ \textit{Phys.Rev.Lett}. \textbf{89}:261101
[arXiv:gr-qc/0207080].

\bibitem{BarCarHaw73} J M Bardeen, B Carter, and S W Hawking (1973) ``The
four laws of black hole mechanics'' \emph{Communications in Mathematical
Physics} \textbf{31}: 161-170

\bibitem{BirDav84} N D Birrell and P C W Davies (1984) \emph{Quantum Fields
in Curved Space} (Cambridge: Cambridge University Press)

\bibitem{Boy69} R H Boyer (1969) ``Geodesic Killing Orbits and Bifurcate
Killing Horizons'' \emph{Proc. R. Soc. Lond}. \textbf{A 311}: 245-252

\bibitem{Cli08} T Clifton (2008)\ \textquotedblleft Properties of Black Hole
Radiation From Tunnelling\textquotedblright\
\emph{Class.Quant.Grav}. \textbf{25}:175022 [arXiv:0804.2635]

\bibitem{Dav76} P C W Davies (1976) ``On the Origin of Black Hole Evaporation
Radiation'' \emph{Proc Roy Soc London} \textbf{A 351}: 129-139.

\bibitem{DavFulUnr76} P C W Davies, S A
Fulling, and W G Unruh (1976) ``Energy-momentum tensor near an
evaporating black hole'' Phys Rev D13: 2720-2723.

\bibitem{Dreetal03} O Dreyer, B Krishnan, E Schnetter, and D Shoemaker
(2003) ``Introduction to Isolated Horizons in Numerical Relativity'' \emph{%
Phys.Rev}.\textbf{D67}:024018 [arXiv:gr-qc/0206008].

\bibitem{Ell84} G F R Ellis (1984) \textquotedblleft Relativistic cosmology:
its nature, aims and problems". In \textit{General Relativity and Gravitation%
}, Ed B Bertotti et al (Reidel), 215-288.

\bibitem{Ell02} G F R Ellis (2002) ``Cosmology and Local Physics" \emph{New
Astronomy Reviews} \textbf{46}: 645-658 [gr-qc/0102017].

\bibitem{Ell13a}
G F R Ellis (2013) ``The arrow of time and the nature of spacetime''
\emph{Studies in History and Philosophy of Modern Physics} 44:
242–262.

\bibitem{Ell13b}
George F R Ellis (2013) ``Astrophysical black holes may radiate, but
they do not evaporate'': arXiv:1310.4771.

\bibitem{EllMaaMac12} G F R Ellis, R Maartens and M A H MacCallum, \emph{%
Relativistic Cosmology} (Cambridge University Press)

\bibitem{Gid13} S B Giddings (2013) \textquotedblleft Black holes. quantum
information, and the foundations of phyiscs\textquotedblright\ \emph{Phyiscs
Today} April 2013: 30-35.

\bibitem{Haj87}
P Hajicek (1987) ``Origin of Hawking radiation'' \emph{Phys. Rev}. \textbf{D 36}, 1065–1079.

\bibitem{Haw73} S W Hawking (1975) \textquotedblleft Particle creation by
black holes\textquotedblright\ \emph{Communications in Mathematical} \emph{%
Physics} \textbf{43}: 199-220.

\bibitem{Haw74} S W Hawking (1974) \textquotedblleft Black-hole
explosions\textquotedblright\ \emph{Nature} \textbf{248}, 30--31.

\bibitem{Haw76} S W Hawking (1976) ``Breakdown of predictability in
gravitational collapse'' \emph{Phys. Rev}. \textbf{D 14}, 2460--2473.

\bibitem{Haw05} S W Hawking (2005) \textquotedblleft Information loss in
black holes\textquotedblright : arXiv:hep-th/0507171.

\bibitem{HawEll73} S W Hawking and G\ F\ R\ Ellis (1973)\ \emph{The Large
Scale Structure of Space-Time} (Cambridge: Cambridge University Press).\

\bibitem{HawPen96} S Hawking and R Penrose (1996) \textit{The Nature of
Space and Time }(Princeton:\ Princeton University Press).

\bibitem{Hen99} R C Henry (1999) \textquotedblleft Kretschmann Scalar for a
Kerr-Newman Black Hole\textquotedblright\ \emph{The Astrophysical Journal}
\textbf{535}: 350--353. [arXiv:astro-ph/9912320].

\bibitem{HorMal04} G T Horowitz and J Maldacena (2004) \textquotedblleft The
black hole final state\textquotedblright . \textit{JHEP} \textbf{0402}:008 [
arXiv:hep-th/0310281].

\bibitem{Mat12} S Mathur (2012) ``The information paradox: conflicts and
resolutions'': arXiv:1201.2079.

\bibitem{Nie08} A B Nielsen (2009) \textquotedblleft Black holes and black
hole thermodynamics without event horizons\textquotedblright\ \emph{%
Gen.Rel.Grav}. \textbf{41}:1539-1584 [arXiv:0809.3850]

\bibitem{Pag76} D N\ Page (1976) \textquotedblleft Particle emission rates
from a black hole: Massless particles from an uncharged, nonrotating
hole\textquotedblright\ \textit{Phys. Rev}.\textbf{\ D 13},
198--206.

\bibitem{ParPqad09} A Paranjape and T Padmanabhan (2009)\ \textquotedblleft
Radiation from collapsing shells, semiclassical backreaction and black hole
formation\textquotedblright\ \emph{Phys.Rev}. \textbf{D80}:044011
[arXiv:0906.1768]

\bibitem{ParWil00} M K Parikh and F Wilczek (2000)\ \textquotedblleft
Hawking Radiation As Tunneling\textquotedblright\ \textit{Phys Rev
Lett} \textbf{85}: 5042-5045.

\bibitem{Pel09} A Peltola (2009) ``Local Approach to Hawking Radiation''
\emph{Class.Quant.Grav}.\textbf{26}: 035014 [arXiv:0807.3309]

\bibitem{Pen65} R Penrose (1965)\ \textquotedblleft Gravitational collapse
and space-time singularities\textquotedblright\ \emph{Phys. Rev.
Lett}. \textbf{14}, 57--59.

\bibitem{PetUza09} P Peter and J-P Uzan (2009) \emph{Primordial Cosmology}
(Oxford: Oxford University Press).

\bibitem{Sch07} M A Schlosshauer (2007), \textquotedblleft Decoherence Is
Everywhere: Localization Due to Environmental Scattering\textquotedblright .
Chapter 3 in \emph{Decoherence and the Quantum-To-Classical Transition},
Frontiers Collection (Heidelberg: Springer): 115-151.

\bibitem{Vis01} M Visser (2003) \textquotedblleft Essential and inessential
features of Hawking radiation\textquotedblright\
\emph{Int.J.Mod.Phys}. \textbf{D1}2 : 649-661
[arXiv:hep-th/0106111].

\end{thebibliography}
\end{document}